# PERFORMANCE ANALYSIS OF HYBRID FORECASTING MODEL IN STOCK MARKET FORECASTING

Mahesh S. Khadka*, K. M. George, N. Park and J. B. Kim[a]

Department of Computer Science, Oklahoma State University, Stillwater, OK 74078, USA
[a]Department of Economics and Legal Studies in Business, Oklahoma State University, Stillwater, OK 74078, USA

`{mahessk, kmg, npark}@cs.okstate.edu, jb.kim@okstate.edu`

## ABSTRACT

*This paper presents performance analysis of hybrid model comprise of concordance and Genetic Programming (GP) to forecast financial market with some existing models. This scheme can be used for in depth analysis of stock market. Different measures of concordances such as Kendall's Tau, Gini's Mean Difference, Spearman's Rho, and weak interpretation of concordance are used to search for the pattern in past that look similar to present. Genetic Programming is then used to match the past trend to present trend as close as possible. Then Genetic Program estimates what will happen next based on what had happened next. The concept is validated using financial time series data (S&P 500 and NASDAQ indices) as sample data sets. The forecasted result is then compared with standard ARIMA model and other model to analyse its performance.*

## KEYWORDS

*Genetic Programming, Concordance, ARIMA, Stock Market Forecasting, Kendall's Tau, Gini's Mean Difference, Spearman's Rho*

## 1. INTRODUCTION

In present financial world, Stock Market forecasting is considered as one of the most challenging tasks. So a lot of attention has been given to analyse and forecast future values and behaviour of financial time series. Different factors interact in stock market such as business cycles, interest rates, monitory policies, general economic conditions, traders' expectations, political events, etc. According to academic investigations, movements in market prices are not random rather they behave in a highly non-linear, dynamic manner [2]. Ability to predict direction and correct value of future stock market values is the most important factor in financial market to make money. These days because of online trading, stock market has become one of the hot targets where anyone can earn profits. So forecasting the correct value and behaviour of stock market has become the area of interest. However, because of high volatility of underlying laws behind the financial time series, it is not any easy task to build such a forecasting model [3].

As mentioned earlier, stock market forecasting has been one of the hot topics among researchers

---

\* **Corresponding Author:** *Mahesh S. Khadka, 219 MSCS, Oklahoma State University, Stillwater, OK 74078, US.A*

*Email address: mahessk@cs.okstate.edu*

over the years. As a result, a lot of researches are conducted and many forecasting model have been proposed. The most common approach taken so far is to use artificial neural networks (ANNs). But using genetic programming in this field is pretty new concept as most studies showed that ANN has some limitations in learning the patterns because stock market data has tremendous noise and complex dimensionality [1]. Moreover, ANN has preeminent learning ability while it is often confronted with inconsistent and unpredictable performance for noisy data. In addition, sometimes the amount of data is so large that the learning of pattern may not work as well. In particular, the existence of continuous data and large amount of data may pose a challenging task to explicit concepts extraction from raw data due to huge amount of data space determined by continuous features [4]. So in this paper, we have presented a hybrid approach based on concordance and genetic programming to predict time series in short term in the same or another time series. The measures of concordance such as Kendall's Tau, Gini's Mean Difference, Spearman's Rho, and a weak interpretation of concordance are used to identify these generic trends in time series. Existing financial data such as S&P 500 and NASDAQ indices are used as sample data sets to validate the concept. The performance of this model is then analysed by comparing with standard ARIMA model and other model proposed by another researcher.

## 2. GENETIC PROGRAMMING

Genetic Programming is a branch of genetic algorithms. The difference between them is the way of representing the solution. Genetic programming creates computer programs as the solution whereas genetic algorithm creates a string of numbers that represent the solution. Here one dimensional vector is called chromosome with element in it is gene. The pool of chromosome is called population.

Genetic Programming uses these steps to solve problems.

  i. Generate a population of random polynomials.

  ii. Compute the fitness value of each polynomial in the population based on how well it can solve the problem.

  iii. Sort each polynomial based on its fitness value and select the better one.

  iv. Apply reproduction to create new children.

  v. Generate new population with new children and current population.

  vi. Repeat step ii – vi until the system does not improve anymore.

The final result that we obtain will be the best program generated during the search. We have discussed how these steps are implemented in our work in next subsections.

### 2.1. Initial Population Generation

The initial population is made of randomly generate programs. We have used traditional grow method of tree construction to construct initial population. A node can be a terminal (value) or function (set of functions +, -, x, /, exp) or variable. If a node is a terminal, a random value is generated. If node is a function, then that node has its own children. This is how a tree grows.

### 2.2. Fitness Evaluation

After initial random population is generated, individuals need to be assessed for their fitness. This is problem specific issue that has to answer "how good or bad is this individual?" In our case, fitness is computed by $\sum_{k=1}^{l}(g(p_k)-f_k)^2$ where *k* is the day in past, *p* is the past data, *f* is the present data and *l* is the length of the section found by concordance measures.

## 2.3. Crossover and Mutation

In crossover, two solutions are combined to generate two new off springs. Parents for crossover are selected from the population based on the fitness of solutions. Mutation is a unary operator aimed to generate diversity in a population and is done by applying random modifications. A randomly chosen subtree is replaced by randomly generated subtree. First, a random node is chose in the tree, and then the node as well as subtree below it is then replaced by a new randomly generated subtree.

## 3. METHODOLOGY

The daily changes for market are well fitted by non-Gaussian stable probability density, which is essentially symmetric with location parameter zero. The time evolution of standard deviation of daily change of stock market follows power law [5]. The Box-Jenkins model requires data to be stationary. Then seasonality has to be checked. Once stationary and seasonality is addressed, then only identification of order of the autoregressive and moving average terms takes place. Same is true with ARIMA model also. The correlation immune to whether biased or unbiased versions for estimation of the variance are used, concordance is not. In this section, we discuss about the forecasting methodology.

### 3.1. Concordance and GP based hybrid model

As we discussed earlier, past data is huge and we want to limit the past data so as to compare with the present using mathematical concordance. Tau, Gini, and Rho concordances of all the possible past segments are compared over a short period of time. This will come out with all the lengths and positions for high concordances. Higher the concordances and longer the matches, indicate better matches. A high concordance means that the trend is likely to continue, so we can use the past data to predict future. To make the prediction as accurate as possible, we search the mathematical equation *g(x)* to map the past data to present data to select which section of the past to use based on the concordances. The genetic program will then search for an equation such that $\forall k, g(p_k) \approx f_k$ where *k* is a day in past. Specifically, we want to minimize. $\sum(g(p_k)-f_k)^2$ for all *k* by choosing the best possible function *g(x)*. The square makes larger differences matter much more than smaller differences. The function *g(x)* will get us close, but it will not be perfect. So we measure the error $e_k$ for each term and subtract that error to get a perfect function. By extrapolating that error and using known values from the past, we can guess values that have not happened yet. This is done through genetic programming.

### 3.2. Forecasting Algorithm

The first step is to find out the pattern from past that looks similar to the present pattern. The algorithm to perform this step is given below.

1. Get stock data for all stocks we want to test.

2. Search for the pattern in the past that look very similar to the present pattern using *Kendall's Tau, Gini's Mean Difference* and *Spearman's Rho* as probabilistic distance measure.

3. Find the highest recorded *Tau concordance* among of all matches.
4. Use *Genetic Program* to match the past trend to present trend as close as possible. Use this program to estimate what will happen next *"now"* based on what happened next *"then"*.
5. Repeat Steps 3 and 4 with *Gini* and *Rho Concordances*.

After the highest value of concordance is obtained, then genetic program should run to find out the best possible solution. The steps involved in genetic programming are as follows.

1. Generate a population of random polynomials *g(x)*.

2. Compute a *"fitness"* of each polynomial, defined by $\sum_{k=1}^{l}(g(p_k)-f_k)^2$ where *g* is the genetic polynomial, *p* is the past data, *f* is the present data, and *l* is the length of the section found by the concordance measures.

3. Sort the polynomials according to their fitness. Then replace the lower half of the population through breeding and mutating the upper half, along with adding new random individuals.

4. Repeat Steps 2 and 3 until a sufficiently low fitness is attained.

## 4. COMPARISON OF MODELS

In this section, we compare the forecasted values obtained from the hybrid model with other models. To test the efficiency of proposed hybrid method, we have used stock index values for S&P 500 and NASDAQ indices values from yahoo finance (http://www.finance.yahoo.com) [6].

### 4.1. Experimental Setup

Here, we have experimented with four cases. The cases are to compare the values obtained from hybrid model and ARIMA model with actual S&P 500 and NASDAQ values for a week (5 business days), for two weeks (10 business days), for three weeks (15 business days) and for four weeks (20 business days) (Data are available upon request).

### 4.2. Result

Table 1 shows the comparison of forecasted values of hybrid model with ARIMA model. In Table 1 and 2, the first column is the test scenarios, the second column shows how many times the predicted direction for both the models are not same as the actual direction of corresponding indices values, the other two columns show the Mean Absolute Percentage Error (MAPE) and Root Mean Squared Error (RMSE) of both the models for each test case scenario.

Table 1. Forecast accuracy comparison for S&P 500 Index.

| Cases | Number of different direction of the model compared to actual direction | | MAPE | | RMSE | |
|---|---|---|---|---|---|---|
| | ARIMA | Hybrid | ARIMA | Hybrid | ARIMA | Hybrid |
| 1 Week | 2 | 0 | 1.14803 | 0.389445 | 18.16397 | 5.31105 |
| 2 Weeks | 5 | 0 | 1.20857 | 0.28719 | 18.25273 | 4.33859 |
| 3 Weeks | 8 | 1 | 1.51134 | 0.54986 | 22.35189 | 9.70717 |

| 4 Weeks | 9 | 3 | 1.91395 | 0.89861 | 28.41956 | 15.90262 |

Table 2. Forecast accuracy comparison for NASDAQ Index.

| Cases | Number of different direction of the model compared to actual direction | | MAPE | | RMSE | |
|---|---|---|---|---|---|---|
| | ARIMA | Hybrid | ARIMA | Hybrid | ARIMA | Hybrid |
| 1 Week | 1 | 0 | 1.38318 | 0.54068 | 42.29495 | 17.29499 |
| 2 Weeks | 5 | 1 | 1.09242 | 0.67182 | 35.66758 | 22.86794 |
| 3 Weeks | 7 | 2 | 1.58685 | 1.08994 | 51.66644 | 41.17626 |
| 4 Weeks | 9 | 3 | 2.13061 | 1.16951 | 69.22673 | 42.22517 |

The result obtained from hybrid model is compared with ARIMA model to show this method is better than existing standard model. In order to do this, parameters, $p$ and $q$, required for ARIMA model are obtained based Bayesian Information Criterion (BIC). Then the values are predicted using ARIMA model.

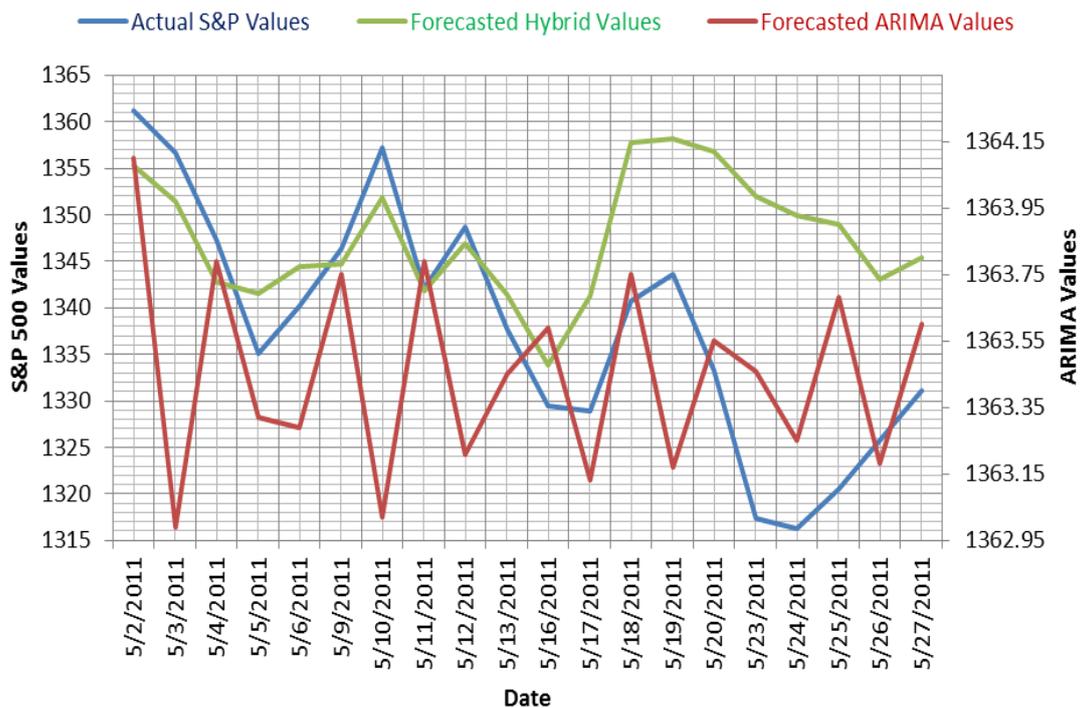

Figure 1. Comparison between actual S&P 500, hybrid model and ARIMA values

Figure 1 shows the graphical representation of actual S&P values, forecasted hybrid model values and forecasted ARIMA model values. Similarly, Figure 2 shows the graphical representation of actual NASDAQ, forecasted hybrid model values and forecasted ARIMA model values. From these figures, it is clear that the ARIMA values are more dispersed from actual values compared to that of hybrid model values. It can also be seen that the values obtained from the hybrid model seems to have almost the same pattern as that of actual values. This also draws the conclusion that the hybrid model performs better than the ARIMA model.

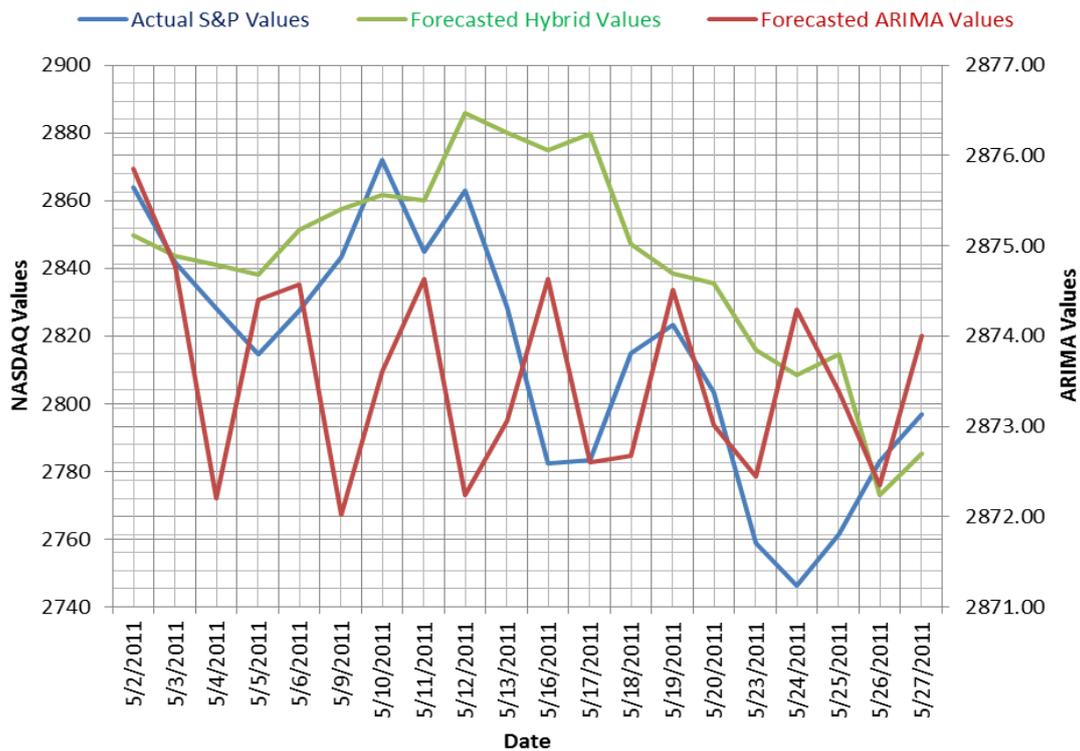

Figure 2. Comparison between actual NASDAQ, hybrid model and ARIMA values

We also calculated MAPE and RMSE in all four cases for both models. Both MAPE and RMSE for hybrid model seem to be better than that of ARIMA model. We have also observed how many times the predicted values do not follow the movement of actual index values in all four cases and calculated efficiency of each model. The efficiency of the model is based on the movement of actual index values turn out to be about on average 80% for hybrid model and about 55% for ARIMA model. From these experimental results, we can clearly say that the hybrid forecasting model performs better than the traditional ARIMA model. Statistical testing is also performed to find out which model performs better with both time series using Wilcoxon Rank-Sum hypothesis test. This test also shows that hybrid model is better than ARIMA model on both data sets.

We have also compared hybrid model results with the results obtained from the model proposed by [3]. According to [3], to test the efficacy of the model, the daily stock price of Apple Computer Inc., IBM Corporation and Dell Inc. are collected from www.finance.yahoo.com. Test data for all three stocks is from 13 September 2004 to 21 January 2005 (91 sequential dataset). Table 3 shows comparison of performance of our hybrid model and forecasting model proposed by [3]. All the data used for [3] are taken from the paper exactly as it is.

Table 3. MAPE Comparison of hybrid model and model proposed by [3]

| Stock Name | MAPE in forecast for 91 sequential test dataset | |
|---|---|---|
| | **Proposed Hybrid Model** | **Model proposed by [3]** |
| Apple Computer Inc. | 1.771113 | 1.9247 |
| IBM Corporation | 0.789556 | 0.84871 |
| Dell Inc. | 0.644035 | 0.699246 |

Paper [3] has considered MAPE as the performance measuring standard and concludes that the method proposed by it is better as it has lower MAPE for the considered stocks than ARIMA model. Based on the same ground of standard, from table 3, we can see that our method has lower MAPE for all three stocks considered by [3]. So, from this experimental result and the data presented by [3], we can say that our hybrid model performs better.

## 5. CONCLUSIONS

Here in this paper, we have analysed the performance of hybrid forecasting model based on concordance and genetic programming. Generally, a comparison between the original time series and a model provides a measure of the model's ability to explain variability in the original time series. Previous studies tried to optimize controlling parameters using global search algorithms. Some focus on the optimization of learning algorithms itself, but most studies had little interest in the elimination of irreverent patterns. This paper has proposed a new hybrid model using genetic programming to mitigate above limitations. This paper not only comes out with a model of forecasting but also compares the result with existing standard ARIMA model and other proposed model and provides enough evidence why this method performs better than those models. The model also turns out to be more consistent which is supported by the fact obtained from experimental results that as the forecasting horizon increases, error level also increases. This hybrid method performs much better in short forecasting horizon. The case that hybrid model performs better is solidified by the conclusion obtained from statistical testing as well.

## REFERENCES

[1]    Han, Ingoo & Kim, Kyoung-jae, (2000) "Genetic algorithms Approach to feature Discretization in artificial neural networks for the prediction of stock price index", *Expert Systems with Applications*, Vol. 10, No. 5, pp120-122.

[2]    Choudhary, Rohit & Garg, Kumkum, (2008) "A Hybrid Machine Learning System for Stock Market Forecasting", *World Academy of Science, Engineering and Technology,* Vol. 39*,* pp315-318.

[3]    Hassan, Md. Rafiul, Nath, Baikunth & Kirley, Micheal, (2007) "A fusion model of HMM, ANN and GA for Stock Market Forecasting", *Expert Systems with Applications,* Vol. 33, pp171-180.

[4]    Liu, Huan & Setiono, Rudy, (1996) "Dimensionality reduction via discretization", *Knowledge-Based Systems,* Vol. 9(1)*,* pp67-72.

[5]    Lan, Boon Leong & Tan, Ying Oon, (2007) "Statistical Properties of Stock Market Indices of different Economies", *Physica A: Statistical Mechanics and its Applications*, *Vol. 375*, No. 2, pp605-611.

[6]    Yahoo Finance Website Historical Prices, http://finance.yahoo.com/, accessed on 29 April 2011.

**Authors**

**Mahesh S. Khadka** received his Bachelor's Degree in Computer Engineering from Nepal Engineering College, Pokhara University, Kathmandu Nepal in 2004 and Master's Degree in Computer Science from Oklahoma State University, Stillwater, Oklahoma, USA in 2008. He is currently a Ph. D student in Computer Science Department at Oklahoma State University. His research interests are in the field of Reliability Theory, Risk Analysis, Simulation and Modeling, and Database.

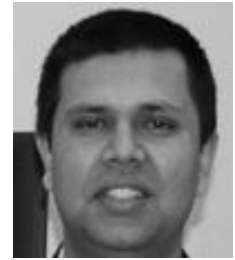

**K. M. George** is a professor in Computer Science Department at Oklahoma State University (one of the major research universities in Oklahoma). He received Ph.D. in Mathematics from the State University of New York at Stony Brook in 1976. During his tenure at Oklahoma State University, he has co-authored more than 100 research papers and given several invited talks. His publications cover several areas of Computer Science – Programming Languages, Expert Systems, Pattern Matching, Mathematical Modeling and Simulation, Risk Analysis, Stock Market Forecasting, and Computer Architecture. As a professor, his major contributions related to DoD projects include G050 system implementation and ABDR support system (ABDRSS) implementation. The G050 system is a major analysis tool used by weapon system managers and government contractors at Tinker and at other ALCs. The ABDRSS was used by the ABDR team at Tinker ALC. As part of summer research supported by Center for Aircraft and Systems/Support Infrastructure (CASI), he conducted research in WUC-NSN-Part number cross-reference, failure mode data collection for indentured parts, and AFMC form 173 automation.

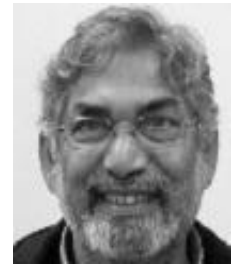

**N. Park** received B. S. and M.S. degrees in computer science from Seoul National University, Seoul, Korea in 1987 and 1989, respectively, and the Ph.D. degree from Texas A&M University, College Station, in 1997. Since 1999, he has been an Associate Professor with the Department of Computer Science, Oklahoma State University, Stillwater, Oklahoma, USA. His research interests include testing and quality assurance of digital systems, design for reliability, risk modeling and analysis computer architecture, defect- and fault-tolerant systems, parallel and distributed computer systems. Dr. Park has been involved in a few research projects funded through Tinker Air Force Base such as: "Engine Data Acquisition, Forecasting, Functionality, and Analysis Module"; "Automated To Management System"; "Evaluate Deficiency Analysis Information Report (DRAIR) Programs for Quality and Accuracy"; as well as a project on "Yield Assurance and Optimization for Clockless Wave pipeline" funded by NSF, during the past ten years..

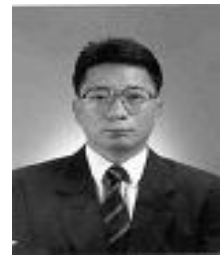

**J. B Kim** is currently an Associate Professor of Economics at Oklahoma State University where he has been teaching and conducting research since 2005. Before that he was a Visiting Assistant Professor and an Assistant Professor at SUNY at Binghamton and at University of St. Thomas, Minnesota. He received his Ph.D. in Economics from the Ohio State University in 2000. His primary research interests are in the areas of international macroeconomics, applied time series econometrics, and empirical financial economics.

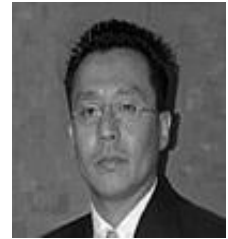